\documentclass[12pt]{iopart}
\pdfoutput=1
\expandafter\let\csname equation*\endcsname\relax
\expandafter\let\csname endequation*\endcsname\relax
\usepackage[numbers,sort&compress]{natbib}
\usepackage{amsmath,amssymb,eucal,graphicx,bm}
\usepackage{subfigure}
\usepackage{float}
\def\ltwid{\mathrel{\raise.3ex\hbox{$<$\kern-.75em\lower1ex\hbox{$\sim$}}}}
\def\gtwid{\mathrel{\raise.3ex\hbox{$>$\kern-.75em\lower1ex\hbox{$\sim$}}}}

\usepackage{appendix}
\usepackage[breakwords]{truncate}
\usepackage{lastpage}
\usepackage{hyperref}
\usepackage{amsfonts}
\usepackage{physics}
\usepackage[section]{placeins}
\usepackage{color}
\usepackage{url}

\usepackage{breakurl}

\begin{document}
\title{When Will an Elevator Arrive?}

\author{Zhijie Feng}
\address{Department of Physics, Boston University, Boston, MA 02215, USA}
\address{Department of Physics, The Hong Kong University of Science and Technology, Clear Water Bay, Hong Kong, China}

\author{S. Redner}
\address{Santa Fe Institute, 1399 Hyde Park Road, Santa Fe, NM 87501, USA}

\begin{abstract}
 
  We present and analyze a minimalist model for the vertical transport of
  people in a tall building by elevators.  We focus on start-of-day
  operation in which people arrive at the ground floor of the building at a
  fixed rate.  When an elevator arrives on the ground floor, passengers enter
  until the elevator capacity is reached, and then they are transported to
  their destination floors.  We determine the distribution of times that each
  person waits until an elevator arrives, the number of people waiting for
  elevators, and transition to synchrony for multiple elevators when the
  arrival rate of people is sufficiently large.  We validate many of our
  predictions by event-driven simulations.

\end{abstract}
  
\section{Introduction} 

How long until the next elevator arrives?  Many of us ponder this question as
we wait in the lobby of a tall building before getting to our destination
floor.  The impact of waiting for elevators is increasing because of the
continued expansion of cities and high-rise buildings.  In Tokyo, New York,
and Hong Kong, for example, there are currently about 190,000~\cite{Tokyo},
84,000~\cite{nyc}, and 69,000~\cite{HK} elevators, respectively.  In the
extraordinarily vertical city of Hong Kong, their number is increasing at a
rate of about 1500 per year~\cite{HK}.  Thus the carrying capabilities of
building elevators necessarily represents an important feature of building
design.

Despite the considerable development of electric elevators since their
inception in the 1880s, as well as their increasing importance in
contemporary society, our understanding of the transport properties of
elevators is incomplete.  There has been much work from the engineering and
operations research perspectives on elevators, including their control and
scheduling in tall buildings (see, e.g.,
\cite{pepyne1997optimal,hikihara1997emergent,schlemmer2002computational,bertsekas1976dynamic,bartz2005validation,siikonen1993elevator,lee2009performance,barney2015elevator,buildings5031070}).
Studies of this genre typically focus either on simulations of realistic
scenarios or on the control mechanisms for multi-elevator systems.  However,
such investigations do not provide insights on the performance of such
systems as a function of basic parameters, such as the passenger demand, as
well as elevator and building characteristics.  The physics-based literature
on the dynamics of elevators has been either primarily numerical in
character~\cite{poschel1994synchronization} or invokes analogies to dynamical
systems theory~\cite{nagatani2003complex,nagatani2004dynamical}.

In this work, we present a simple-minded probabilistic approach to treat a
demand-driven elevator system and develop insights about the performance of
such a system.  We focus on the start of a workday, in which people enter a
building lobby at a given rate and want to get to their destination floors.
This scenario is sufficiently simple that some analytical results can be
obtained, yet this case still reflects realistic aspects of elevator
operation.  Within a minimal model to be defined in the next section, we
determine the distribution of times that one has to wait for an elevator and
its dependence on the arrival rate of individuals, the number of elevators,
and the capacity of each elevator.  We validate many of our predictions by
event-driven simulations.  We also examine the conditions under which
multiple elevators tend to synchronize.  This latter property bears some
resemblance to the clustering phenomenon that occurs in subways and along bus
routes~\cite{o1998jamming}, where many full vehicles arrive in quick
succession at a subway station or a bus stop, followed by a long period with
no vehicles arriving.

In the next section, we outline our model.  In Sec.~\ref{sec:inf}, we treat
the dynamics in the simplifying case of a single infinite-capacity elevator.
We derive the time of a single elevator cycle and its distribution, as well
as the distribution of the number of passengers in the elevator.  We then
turn to the case of a single finite-capacity elevator in
Sec.~\ref{sec:finite}, where we first discuss the condition for a steady
state, and then present basic dynamical properties, such as the ``clearing''
time---the time interval between events where all waiting passengers are
accommodated in the elevator that is currently loading---and the clearing
probability, as well as the occupancy distributions in the lobby and in the
elevator.  In Sec.~\ref{sec:k}, we treat the realistic situation of many
finite-capacity elevators.  We determine the steady-state condition and then
investigate how synchronization can occur.  Some concluding remarks are given
in Sec.~\ref{sec:concl}

\section{Model}

\begin{figure}[ht]
  \centerline{\includegraphics[width=0.65\textwidth]{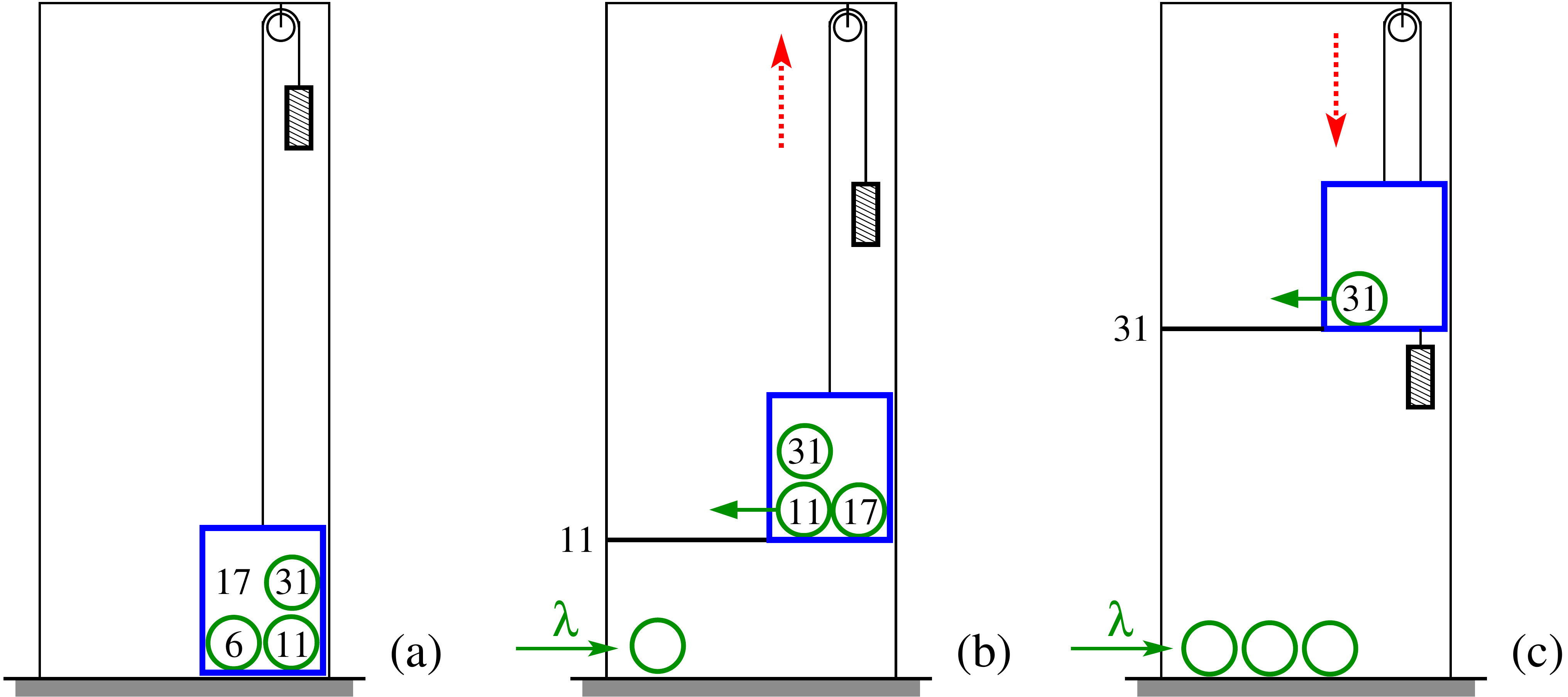}}
  \caption{Cartoon of single-elevator transport during start-of-day operation
    where passengers arrive at rate $\lambda$: (a) Passengers (circles) just
    after entering the elevator; the numbers indicate destination floors. (b)
    A passenger leaves the elevator at her/his destination ($11^{\rm th}$)
    floor.  (c) After the last passenger leaves, the elevator returns to the
    ground floor.}
  \label{fig:cartoon}
\end{figure}

Our model is based on the following assumptions (Fig.~\ref{fig:cartoon}):

\begin{enumerate}

\item Start-of-day operation: the building is initially unoccupied, and
  individuals arrive at the ground floor lobby according to a Poisson process
  at rate $\lambda$. 

\item When an elevator reaches the lobby, it is filled on a
  first-come/first-serve basis until either all passengers are accommodated
  or the elevator reaches its capacity $C$.

\item The building has $F$ floors and $k$ identical elevators that
  can access all floors.

\item Each person has a distinct destination floor that is uniformly
  distributed in $[1,F]$.

\item The time for an elevator to travel one floor is $\tau_e$.

\item Each elevator stop requires a time $\tau_s$ per entering and exiting
  person that is independent of the elevator occupancy.

\end{enumerate}

While most of these features of the model accord with everyday experience,
various approximations have been made and other relevant attributes have been
neglected.  These include: (a) In some tall buildings, some elevators only
stop at a subset of all floors.  This restriction vaguely resembles staging
in a multistage
rocket~\cite{hall1958optimization,doi:10.1080/0020739740050101}, a device
that leads to greater efficiency.  (b) The time to enter and exit an elevator
is not constant, but is clearly an increasing function of its occupancy.  (c)
Most skyscrapers have a smaller floor area in the higher stories, so the
distribution of destination floors is not uniform.  (d) Travel between
different building floors or from an upper floor back to the lobby is not
treated.  Incorporating all these features would be more realistic, but such
a generalization would greatly complicate theoretical modeling.  For both
parsimony and tractability, we only include the elements (i)--(vi) listed
above.  Another desirable feature of this minimalist model is that it can be
simulated with great efficiency by an event-driven approach (see
\ref{app:alg} for details).

\section{Single Infinite-Capacity Elevator}
\label{sec:inf}

We first investigate the idealized case of a building with a single
unlimited-capacity elevator.  While patently unrealistic, this situation
provides the starting point for treating finite-capacity elevators and
multi-elevator buildings.  With a single infinite-capacity elevator, a steady
state is eventually achieved in which the average time for the elevator to
complete a single cycle, i.e., return to the ground floor, equals the average
number of people who arrive in the lobby during a cycle.  Note that the
infinite-capacity case is equivalent to the individual arrival rate $\lambda$
being sufficiently small that a finite elevator capacity is never reached.
We now determine basic features of this steady state.

\subsection{The cycle time}

\begin{figure}[ht]
  \centerline{\includegraphics[width=0.5\textwidth]{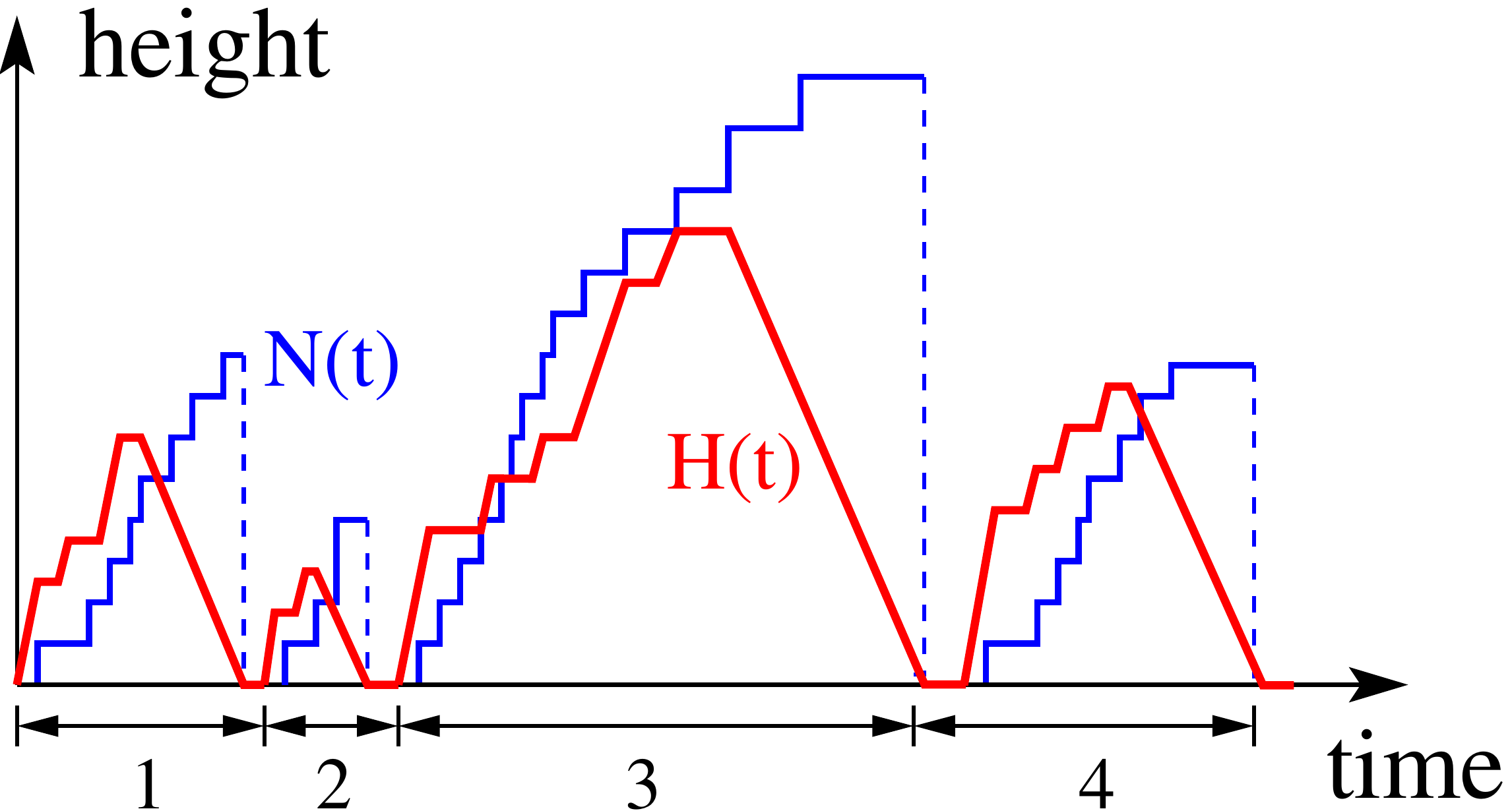}}
  \caption{Schematic time dependence of the number of passengers, $N(T)$, in
    the lobby (blue) and the elevator height, $H(t)$ (red).  Four elevator
    cycles are indicated.}
  \label{fig:cycle}
\end{figure}

A single cycle of an elevator involves the following steps
(Figs.~\ref{fig:cartoon} \& \ref{fig:cycle}):
\begin{enumerate}
\item The elevator arrives on the ground (lobby) floor.
\item Waiting passengers in the lobby enter the elevator.
\item The elevator delivers each passenger to her/his destination floor in
  ascending order and passengers with this destination floor exit.
\item When the elevator empties, it returns to the ground floor and a cycle
  begins anew.
\end{enumerate}
We first determine the time for a single elevator cycle when $N$ passengers
have entered the elevator.  This cycle time is obviously an increasing
function of $N$, and two factors contribute to this $N$ dependence.  First,
the total time that the elevator is stopped to pick up and discharge
passengers in a single cycle in $2N\tau_s$.  Second, for increasing $N$, it
is more likely that the elevator goes to a higher floor to discharge the last
passenger with the highest destination floor.  For $N\gg 1$ passengers, we
use extreme-value
statistics~\cite{gumbel2012statistics,galambos1978asymptotic} to find that
the expected highest destination floor among $N\gg 1$ passengers is given by
\begin{align*}
  F_{\rm max} = F\,\frac{N}{N+1}\,.
\end{align*}
Consequently, the expected time for the elevator to complete a single cycle
is
\begin{align}
  \label{TN}
T(N) &= 2F_{\rm max}\, \tau_e+ 2N\tau_s = 2F\tau_e\,\frac{N}{N+1}+ 2N\tau_s\,.
\end{align}

To obtain a rough estimate of the cycle time, we use that $\tau_e=1\,$sec,
$\tau_s=2.5\,$sec, and $F=100$ floors; these are representative numbers for
elevators in a tall building~\cite{wiki-elevator}.  The cycle time is then
$T(N)= 200 N/(N\!+\!1)+5N$.  For $N=20$ passengers, which is typical for a
high-capacity elevator, the expected time for one elevator cycle is
$T(20)\approx 290\,$sec $\approx 5\,$min.  Henceforth, we fix $\tau_e=1$ for
simplicity, so that $\tau_s$ becomes the ratio of the single-passenger
entrance/exit time to the single-floor travel time.

\subsection{The steady state}
\label{subsec:ss}

In a single cycle of duration $T$, $\lambda T$ new passengers typically
arrive after the elevator leaves before is returns.  The steady-state
occupancy of an elevator, $ N_{\rm ss}$, is determined by equating
$\lambda T(N)$ in Eq.~\eqref{TN} with $N$.  This gives
\begin{align*}
\lambda\left[ 2F\Big(\frac{N_{\rm ss}}{N_{\rm ss}+1}\Big)+ 2N_{\rm ss}\tau_s\right]=N_{\rm ss}\,,
\end{align*}
from which
\begin{align}
\label{Nss}
  N_{\rm ss}= \frac{2\lambda F}{1-2\lambda \tau_s}-1\,.
\end{align}
A basic consequence of this simple calculation is the existence of a critical
arrival rate $\lambda_c=1/2\tau_s$.  When the arrival rate exceeds
$\lambda_c$, progressively more passengers will be waiting for the elevator
after each successive cycle and no steady state is possible.

\begin{figure}[ht]
  \centerline{\includegraphics[width=0.45\textwidth]{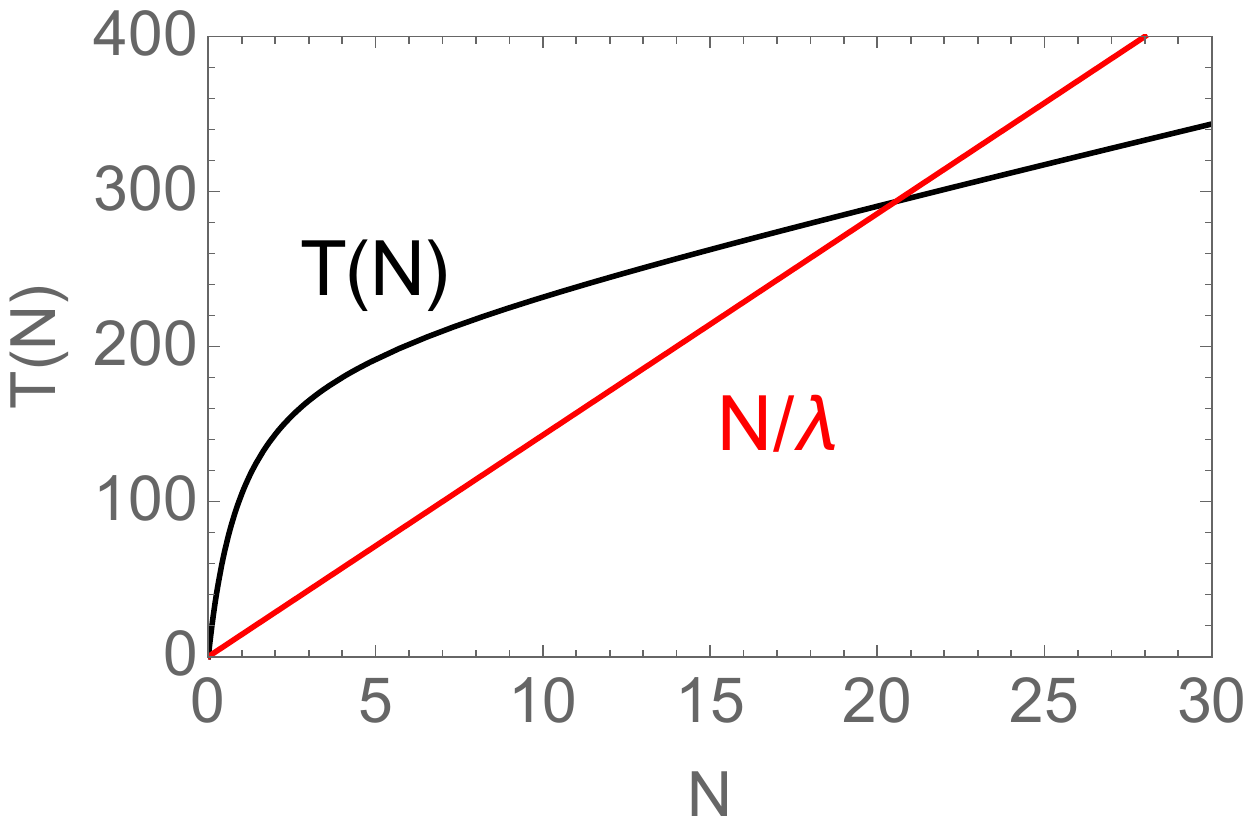}}
  \caption{Dependence of the elevator cycle time $T(N)$ in Eq.~\eqref{TN} on
    $N$ for $\tau_e=1\,$sec, $\tau_s=2.5\,$sec and $F=100$ floors.  A steady
    state arises when $T(N)$ intersects the line $N/\lambda$, which occurs
    for $\lambda<\lambda_c=1/5$ for these parameters.  In this example
    $\lambda=0.07<\lambda_c$}
  \label{fig:NT}
\end{figure}

It may be surprising at first sight that the critical arrival rate does not
depend on the building height.  This independence occurs because of the
infinite elevator capacity and because the travel time, $2NF/(N\!+\!1)$,
becomes a constant contribution to the total cycle time for large $N$, which
becomes negligible for $N\to\infty$.  Consequently, the dependence on
building height disappears.  As a numerical example, for $F=100$ and
$\tau_s=2.5$, $N_{\rm ss}=200\lambda/(1-5\lambda)-1$.  For a steady-state
elevator occupancy of $N_{\rm ss}=20$, $\lambda = 21/305 \approx 0.07$ (the
intersection point in Fig.~\ref{fig:NT}).  Thus this single-elevator system
is not close to its transport limit of $\lambda_c=1/5$ when $N_{\rm ss}=20$.
Since all waiting passengers can fit on the elevator, the longest that any
passenger has to wait is one complete cycle; this occurs when the next
passenger arrives just after the elevator has departed.

\subsection{Cycle and occupancy distributions}

To compute the distribution of cycle times, we use Eq.~\eqref{TN} to write
the maximum floor reached by the elevator in terms of $N$ and the cycle time:
$F_{\rm max} =\frac{1}{2}(T-2N\tau_s)$.  When the distribution of destination
floors is uniformly distributed in $[1,F]$, the probability that the maximum
floor reached by an elevator with $N$ passengers
is~\cite{gumbel2012statistics,galambos1978asymptotic}
\begin{align}
  \label{MF}
  M(F_{\rm max}) = \frac{N}{F}\left[\frac{F_{\rm max}}{F}\right]^{N-1}\,.
\end{align}
Since we are considering tall buildings, the assumption $F\gg 1$ is implicit.

We now use $M(F_{\rm max})\,dF_{\rm max}=P(T)\,dT$, with $P(T)$ defined as
the cycle-time distribution, to eliminate $F_{\rm max}$ in favor of $T$ to
determine $P(T)$.  To compute the distribution of times for the $i^{\rm th}$
cycle, $P_i(T)$, we must sum over the possible values of $N$ in a given cycle.
This leads to
\begin{subequations}
\label{iter}
\begin{align}
  P_i(T) = \sum_{N=1}^{T/2\tau_s}
  \frac{N}{2F}\left[\frac{\frac{1}{2}(T-N\tau_s)}{F}\right]^{N-1} Q_i(N)\,,
\end{align}
where $Q_i(N)$ is the probability that the elevator has $N$ passengers in the
$i^{\rm th}$ cycle.  The upper limit on the sum corresponds to all occupants
of the elevator having the smallest possible destination floor.
Equivalently, this limit corresponds to the maximum number of passengers that
can be accommodated for a given cycle time.

Because new passengers arrive at rate $\lambda$, the probability $Q_i(N)$ is
given by the Poisson distribution that is integrated over all possible values
of the previous cycle time:
\begin{align}
  Q_i(N)= \int dT\, e^{-\lambda T}\, \frac{(\lambda T)^N}{N!} \, P_{i-1}(T)\,.
\end{align}
\end{subequations}  
That is, the number of passengers waiting for the $i^{\rm th}$ cycle of the
elevator depends on the $(i-1)^{\rm st}$ cycle time.  In turn, the parameters
in the $(i-1)^{\rm st}$ cycle depend on the parameters in the
$(i-2)^{\rm nd}$ cycle.  Thus the distributions $P_i(T)$ and $Q_i(N)$ have to
be determined iteratively.

\begin{figure}[ht]
  \center{
    \subfigure[]{\includegraphics[width=0.49\textwidth]{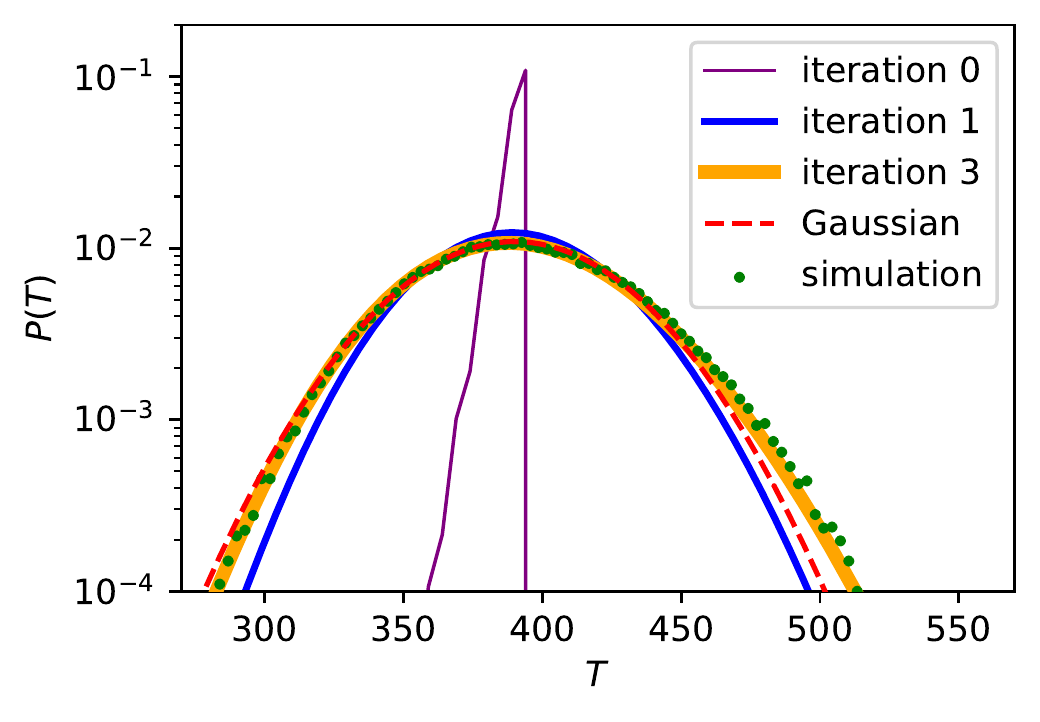}}
     \subfigure[]{\includegraphics[width=0.49\textwidth]{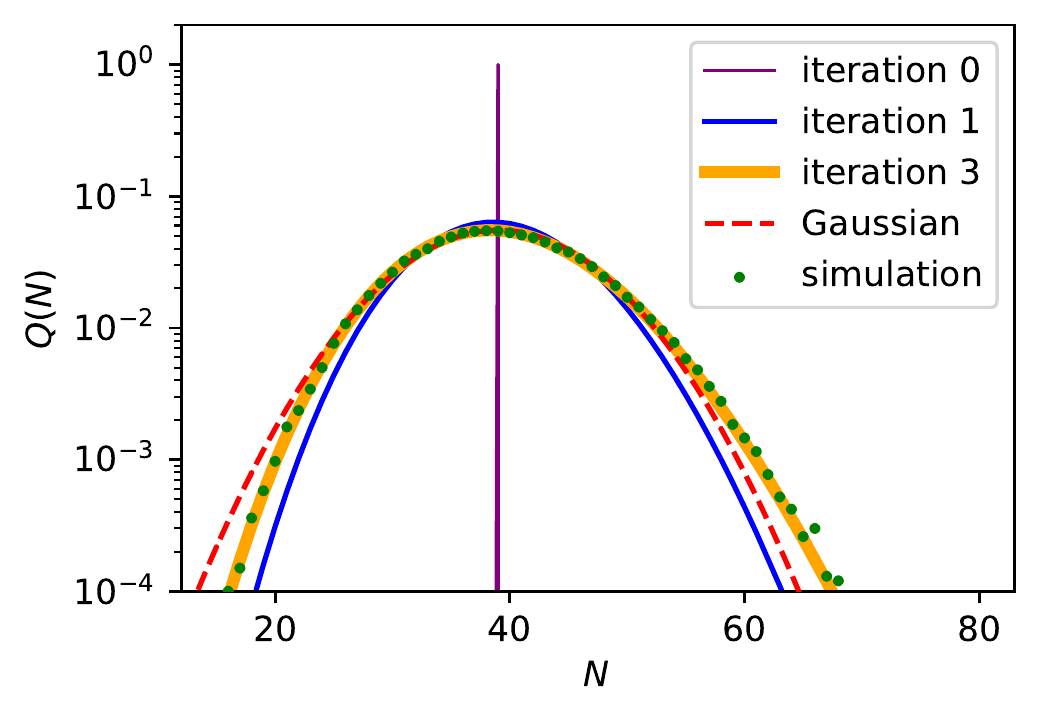}}}
   \caption{(a) The cycle time distribution $P(T)$ (binned to eliminate
     artificial spikes) and (b) the occupancy distribution $Q(N)$ for the
     case of arrival rate $\lambda =\lambda_c$/2, with $\tau_s=2.5$ for all
     passengers.  For rapid convergence, the initial condition is chosen to
     be the steady-state occupancy, $N_0=N_{ss}=39$.}
  \label{fig:T-distribution}
\end{figure}

To illustrate this iterative approach, suppose that initially there are $N_0$
passengers waiting for the elevator.  Then the probability distribution for
the time of this zeroth cycle is
\begin{align*}
  P_0(T) =
  \frac{N_0}{2F}\left[\frac{\frac{1}{2}(T-N_0\tau_s)}{F}\right]^{N_0-1}\,.
\end{align*}
Correspondingly, the probability that $N$ passengers waiting for the elevator
at the start of the first cycle is
\begin{align*}
  Q_1(N) = \int dT\, e^{-\lambda T}\, \frac{(\lambda
  T)^N}{N!} \,  \frac{N_0}{F}\left[\frac{\frac{1}{2}(T-N_0\tau_s)}{F}\right]^{N_0-1}\,.
\end{align*}
Using Eqs.~\eqref{iter}, we iterate to give the distributions of $N$ and $T$
in successive elevator cycles.  Numerically, this iteration quickly converges
to a steady state (Fig.~\ref{fig:T-distribution}).  By making the assumption
of a steady state, we may write closed equations for the distributions $P$
and $Q$ by dropping the subscript and eliminating $Q$ in the equation for $P$
(and vice versa).  We thereby find the implicit solutions:
\begin{align}
\label{PQ}
\begin{split}
  P(T) &= \int_0^{T/2\tau_s} dx\,\, \frac{x}{2F}\left[\frac{\frac{1}{2}(T-x\tau_s)}{F}\right]^{x-1}\int_0^\infty dy\,\, 
  \!e^{-\lambda y}\,\, \frac{(\lambda y)^x}{x!} \,P(y)\,,\\[2mm]
  Q(N) &= \int_0^{\infty} dy\,e^{-\lambda y}\int_0^{T/2\tau_s} dx\,\, \frac{x}{2F}\left[\frac{\frac{1}{2}(T-x\tau_s)}{F}\right]^{x-1}
  \! \frac{(\lambda y)^x}{x!} \,Q(x)\,. 
\end{split}
\end{align}
In our numerical solutions, we found it simpler, however, to iterate
Eqs.~\eqref{iter} to find the steady-state distributions, the results of
which are shown in Fig.~\ref{fig:T-distribution}.

In terms of the cycle time distribution, we can now determine the more
relevant distribution of times that an individual has to wait before an
elevator arrives.  Since an individual arrives equiprobably during an
elevator cycle, her/his waiting time is uniformly distributed in the range
$[0,T]$ for a given elevator cycle time $T$.  We now average this uniform
distribution over all cycle times of duration $t$ or larger to obtain the
individual waiting time distribution, which we define as $\mathcal{W}(t)$:
\begin{align}
\label{waitT}
\mathcal{W}(t)=\int_t^{\infty}\frac{P(T)}{T}\,dT\,.
\end{align}
For the peaked and close-to-Gaussian cycle time distribution in
Fig.~\ref{fig:T-distribution}(a), the waiting time distribution resembles the
Fermi-Dirac distribution at low temperatures---nearly constant for small
times and then rapidly cut off beyond the average cycle time.

\subsection{Fluctuations}

As shown in previous section, the cycle time and occupancy distributions
quickly converge to steady-state forms with well-localized peaks.  We can
determine the widths of these two distributions by using the mean values
$T_{ss}$ and $N_{ss}$ from Sec.~\ref{subsec:ss}, exploiting the Poissonian
nature of the distribution for $N$, and also making use the law of total
variance~\cite{weiss2006course}.

We denote $E(x)=\langle x\rangle$ as the mean value of the variable $x$ and
${\rm Var}(x)=\langle x^2\rangle-\langle x\rangle^2$ as its variance.
Because the distribution of $N$ is Poissonian with mean value $\lambda T$ for
a given value of the travel time $T$, ${\rm Var}(N|T)=\lambda T$.  The law of
total variance states that
${\rm Var}(N)={\rm E}\big({\rm Var}(N|T)\big)+{\rm Var}\big({\rm
  E}(N|T)\big)$~\cite{weiss2006course}.  For the elevator system, this gives
\begin{align}
{\rm Var}(N)&={\rm E}\big({\rm Var}(N|T)\big)+{\rm Var}\big({\rm E}(N|T)\big) \nonumber\\
&=\lambda {\rm E}(T)+{\rm Var}(\lambda T) \nonumber\\
&=N_{ss}+\lambda^2 {\rm Var}(T) \ .
\label{varN}
\end{align}

Since the elevator cycle time equals $T=2F_{\rm max}+ 2N\tau_s$ for a given
value of $N$, the law of total variance for $T$ now leads to
\begin{align}
{\rm Var}(T)&={\rm E}\big({\rm Var}(T|N)\big)+{\rm Var}\big({\rm E}(T|N)\big) \nonumber\\
            &={\rm E}\big({\rm Var}(2F_{\rm max})+2N\tau_s\big)
              +{\rm Var}(2F_{\rm max}+ 2N\tau_s) \nonumber\\
&=4{\rm E}\big({\rm Var}(F_{\rm max})\big) +4\tau_s^2\,{\rm Var}(N) \ . \nonumber
\end{align}
We now use the distribution \eqref{MF} for $F_{\rm max}$ to compute the
$\text{Var}(F_{\rm max})$ and obtain
\begin{align}
  {\rm Var}(F_{\rm max}) ={{N F^2}/\big[{(N+1)^2 (N+2)}\big]}~
    \underset{N\to\infty}{\longrightarrow}~ \left({F}/{N}\right)^2\,.\nonumber
\end{align}
Thus
\begin{equation}
{\rm Var}(T)=4\left({F}/{N_{ss}}\right)^2 +4\tau_s^2\,{\rm Var}(N) \ .
\label{varT}
\end{equation}
Substituting \eqref{varT} in \eqref{varN} and also using \eqref{varT} itself,
we finally have
\begin{align}
\begin{split}
{\rm Var}(T)&=4\left(\frac{F}{N_{ss}}\right)^2 +4\tau_s^2\big[\lambda T_{ss}+\lambda^2 {\rm Var}(T) \big]=\frac{4(F/N_{ss})^2+4\tau_s^2N_{ss}}{1-4\tau_s^2\lambda^2} \ ,\\ 
{\rm
  Var}(N)&=N_{ss}+\frac{4\lambda^2(F/N_{ss})^2+4\lambda^2\tau_s^2N_{ss}}{1-4\tau_s^2\lambda^2}
\ .
\end{split}
\label{varTN}
\end{align}

We now numerically estimate Var$(T)$ and Var$(N)$ and compare with
Fig.~\ref{fig:T-distribution}.  In this figure, $\lambda=\lambda_c/2$, which
leads to $N_{ss}=39$.  The cycle time from \eqref{TN} then is $T(39)=390$.
Using $\tau_s=2.5$ and $F=100$, we obtain ${\rm Var}(T)\approx 1335$ and
${\rm Var}(N)\approx 53$.  Consequently, the waiting time between successive
arrivals of the elevator will typically be in the range 390 sec $\pm$ 36 sec,
while the number of passengers in the elevator will be 39 $\pm$ 7.  These
numbers are in accord with the simulated distributions in
Fig.~\ref{fig:T-distribution}.  Thus fluctuations in the cycle time and
occupancy are substantial, but do not dominate the steady-state behavior.

\section{Single Finite-Capacity Elevator}
\label{sec:finite}

We now turn to the slightly more realistic case of a single elevator with a
finite capacity $C$.  This discussion serves as a starting point to treat
multiple identical elevators.

\subsection{The steady state}
\label{subsec:ss1}

For a single elevator in the steady state, the average number of passengers
waiting when the elevator arrives at the ground floor must be less than or
equal to $C$.  At the stability limit, the elevator should be filled to
capacity.  Thus a steady state should arise whenever $\lambda T(C)\leq C$,
where $T(C)$ is the cycle time for an elevator filled to capacity and
$\lambda$ is again the passenger arrival rate.  Using Eq.~$\eqref{TN}$ for
$T(C)$, we have
\begin{align}
\label{lambdac}
  \lambda \leq \lambda_c= \frac{1}{2\tau_s+2F/(C+1)}\,.
\end{align}
For an elevator of capacity $C=20$ and using $\tau_s=2.5\,$sec, $F=100$, the
steady-state criterion gives $\lambda \leq 0.0689$.  Thus, for a single
elevator, the critical arrival rate is reduced by nearly a factor of 3
compared to an infinite-capacity elevator, where $\lambda_c=1/5$
(Sec.~\ref{subsec:ss}).  A critical rate of $\lambda_c\approx 0.07$
corresponds to an unrealistically small arrival rate of approximately 1
passenger every 15 seconds.  Clearly, and as we all have experienced, many
elevators are needed to service a tall office building, as will be discussed
in Sec.~\ref{sec:k}

\subsection{The clearing time}
\label{subsec:ct}

\begin{figure}[ht]
  \centerline{\includegraphics[width=0.7\textwidth]{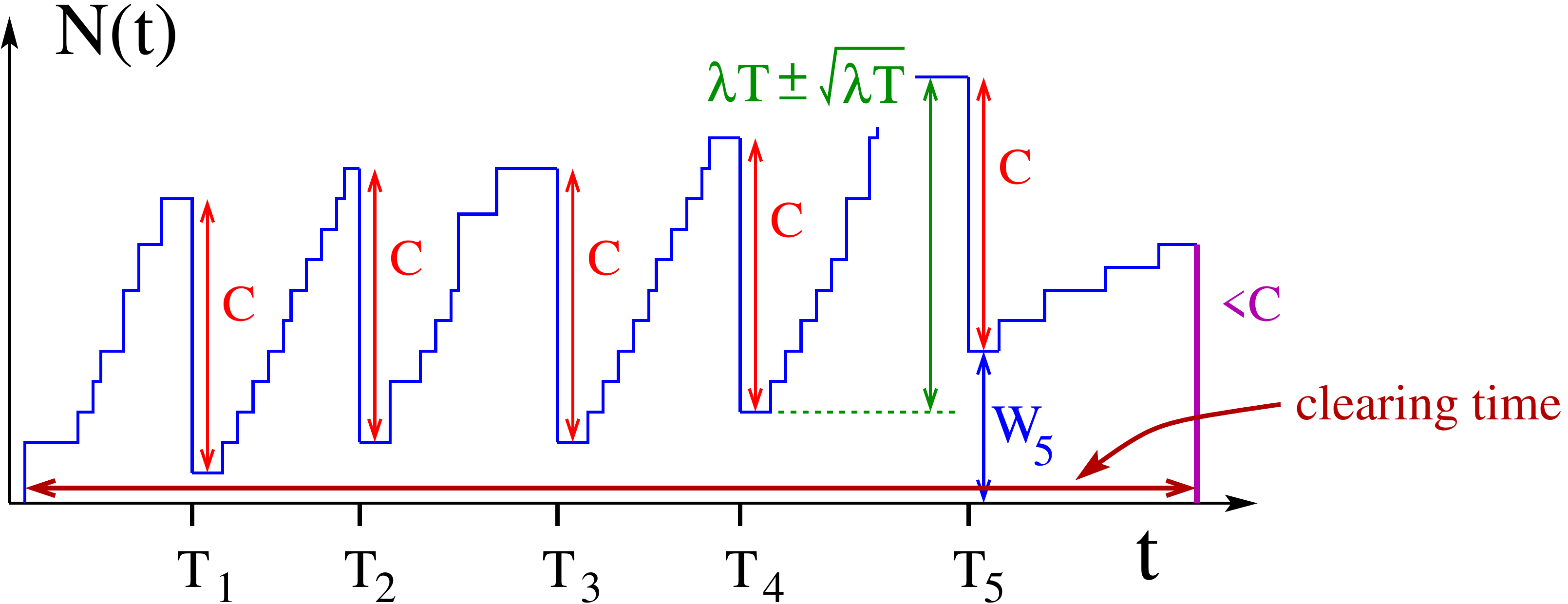}}
  \caption{Schematic illustration of the time dependence of the number of
    people $N(t)$ waiting on the ground floor over six elevator cycles, with
    clearing occurring at the sixth cycle.  The number of new passengers
    accumulated in each cycle is $\lambda T \pm \sqrt{\lambda T}$, while at
    most $C$ passengers are removed in each cycle.}
  \label{fig:fluct}
\end{figure}

A basic characteristic of single-elevator dynamics is the ``clearing'' time,
defined as the time interval between the successive events where the lobby is
emptied when the elevator leaves the ground floor (Fig.~\ref{fig:fluct}).
When the elevator takes on passengers, the number of waiting passengers
$N(t)$ suddenly decreases by $C$, if the number of waiting passengers exceeds
the elevator capacity, or by $N(t)$, otherwise.  For the example shown in the
figure, there is a transient buildup of waiting passengers who have to wait
more than one cycle, under the first come/first serve assumption, before
boarding an elevator.  Eventually, the situation arises in the sixth cycle
where $N(t)<C$, at which point the lobby is cleared when the elevator leaves.

To determine the clearing time, we appeal to an equivalence between the
buildup and removal of passengers in the lobby and a random walk process.
Consider first the case where the arrival rate equals its critical value, for
which $\lambda_c T=C$.  Thus the buildup of passengers in the lobby during
one elevator cycle and the removal of waiting passengers when an elevator is
loaded are, on average, equal.  Let $W_n$ denote the number of people still
waiting in the lobby after the $n^{\rm th}$ cycle.  Note that whenever the
lobby is not cleared, the elevator will be at its capacity of $C$ passengers.
Therefore, the cycle time is almost a constant and is narrowly distributed
about the value $T=2F\tau_e\,\frac{C}{C+1}+ 2C\tau_s$.  For simplicity, in
this and in the following subsection only, we define $T$ as the cycle time of
a full elevator, in which the maximum floor reached by the elevator is
deterministic and equal to its average value of $C/(C+1)$.

\begin{figure}[ht]
  \center{
    \subfigure[]{\includegraphics[width=0.44\textwidth]{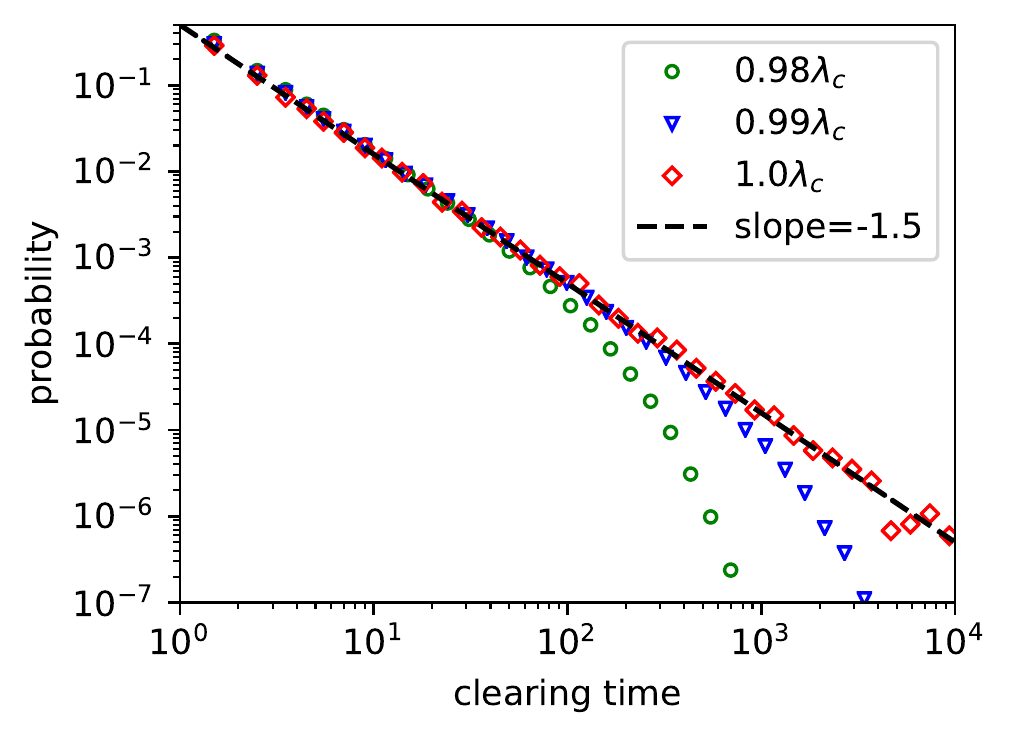}}\qquad
     \subfigure[]{\includegraphics[width=0.41\textwidth]{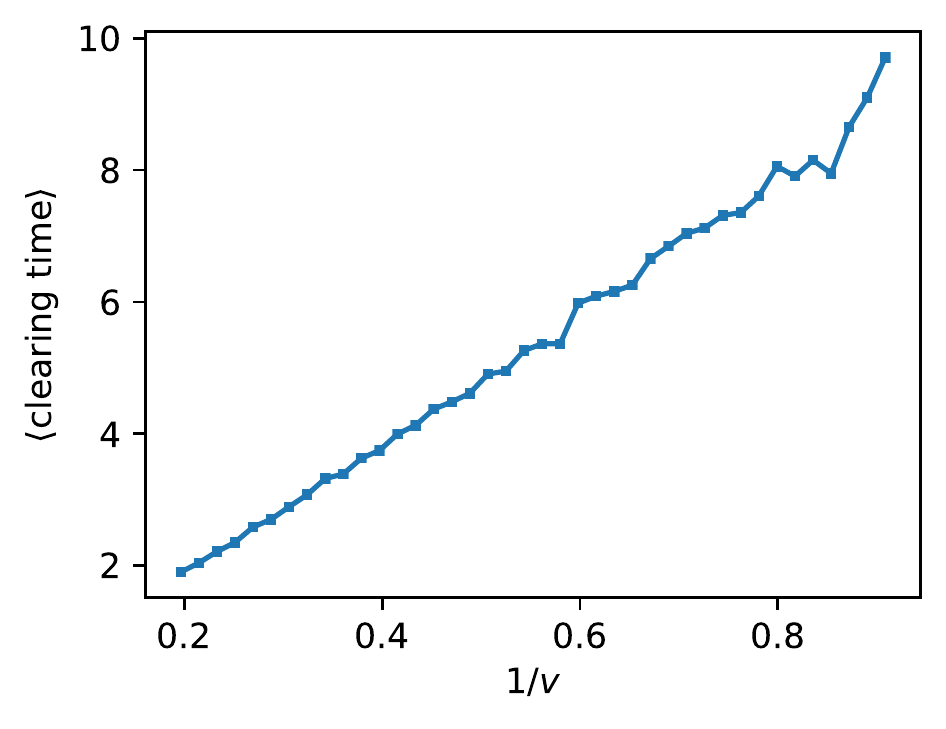}}}
   \caption{Simulation results for: (a) the clearing time probability for
     various $\lambda$ values, and (b) the average clearing time versus
     $1/v$ with $\lambda$ between $0.9\lambda_c$ and $0.98\lambda_c$, for a
     single elevator with capacity 50. }
\label{fig:RTD}
\end{figure}

For $\lambda=\lambda_c$, $W_n$ equiprobably increases or decreases by an
amount of typical magnitude $\sqrt{\lambda T}$ (Fig.~\ref{fig:fluct}).  That
is, $W_n$ undergoes an unbiased random walk of step size $\sqrt{\lambda T}$,
subject to an absorbing boundary condition whenever $W_n=0$. This latter
condition corresponds to the lobby being cleared.  Consequently, by the
connection to the one-dimensional first-passage process, we know that the
average clearing time is infinite and its distribution asymptotically decays
as $n^{-3/2}$~\cite{redner2001guide}, as illustrated in
Fig.~\ref{fig:RTD}(a).

In the realistic case of $\lambda\ltwid \lambda_c$, the average number of
passengers waiting in the lobby will typically decreases by
$C-\lambda T\equiv v$ in each cycle.  We are primarily interested in the case
where this systematic decrease is smaller than the random-walk step size,
$\sqrt{\lambda T}$.  The opposite limit is wasteful from a practical
viewpoint, because it would correspond to a large excess of elevator
capacity.  Thus we treat the limit where the bias $v$ is small but negative
during the start of the day rush.  In this case $W_n$ undergoes a weakly
biased random walk, again with a typical step size of $\sqrt{\lambda T}$.
Now the distribution of clearing times will have the same $n^{-3/2}$ tail as
in the critical case, but with a exponential cutoff due to the bias, which
leads to the average clearing time scaling as $1/v$~\cite{redner2001guide}.
This latter behavior is illustrated in Fig.~\ref{fig:RTD}(b).

\subsection{The clearing probability}
\label{subsec:cp}

In addition to the clearing time, another useful indicator of the efficiency
of the elevator system is the clearing probability $\mathbf{P}$, which we
define as the probability that the lobby is cleared of all waiting passengers
when the elevator leaves.  We compute this probability in the current cycle
in terms of the state of the system in the previous cycle.  There are two
cases that need to be considered: either the lobby was (a) cleared or (b) not
cleared in the previous cycle.  For case (a), we further require that the
number of newly arriving passengers before the elevator next arrives does not
exceed the elevator capacity $C$.  In case (b), the number $W$ of remaining
waiting passengers from last cycle plus the number of newly arriving
passengers cannot exceed $C$.  Under the assumption of a steady state, we
have
\begin{align}
    \label{mathP}
  \mathbf{P}=\mathbf{P}\,\mathcal{Q}(C|{\rm cleared})
  +(1-\mathbf{P})\sum_{W}  R(W)\,\mathcal{Q}(C-W|{\rm not \, cleared})\,.
\end{align}
Here $\mathcal{Q}(X|\cdot)=\sum_N^XQ(N|\cdot)$ in the cumulative conditional
probability that the number of new arriving passengers is $X$ or less, given
that last cycle is either cleared or not cleared.  The first term in
\eqref{mathP} gives the contribution due to case (a) and the second term
accounts for case (b).  In this second term, $R(W)$ is the probability that
$W$ passengers remain in the lobby after the elevator leaves, given that the
lobby was not cleared in the previous cycle.  Subsequently $C-W$ or fewer new
passengers can arrive before the elevator next arrives so that the lobby will
be cleared in the current cycle.

We now determine the factors in \eqref{mathP}.  We compute the stationary
distribution $R(W)$ from the biased random-walk description of
Sec.~\ref{subsec:ct} for the number of people waiting in the lobby.  For this
random walk, the step size is $\delta x=\sqrt{\lambda T}$ and the bias is
$v=C-\lambda T$.  Since the unit of time in this random walk process is one
elevator cycle, the single-step time is $\delta t=1$, from which the
diffusion constant is $D=\delta x^2/(2\delta t)=\lambda T/2$.  The stationary
probability distribution of this biased random walk, subject to a absorbing
boundary condition at $W=0$, is (see, e.g., \cite{crank1979mathematics})
\begin{align*}
  R(W)=\frac{v}{D}\,e^{-vW/D}=2(\lambda_c/\lambda-1)e^{-2(\lambda_c/\lambda-1)W}\,,
\end{align*}
with
$v/D=(C-\lambda T)/(\lambda T/2)=2(\lambda_c/\lambda-1)$.

\begin{figure}[ht]
\centerline{\includegraphics[width=0.45\textwidth]{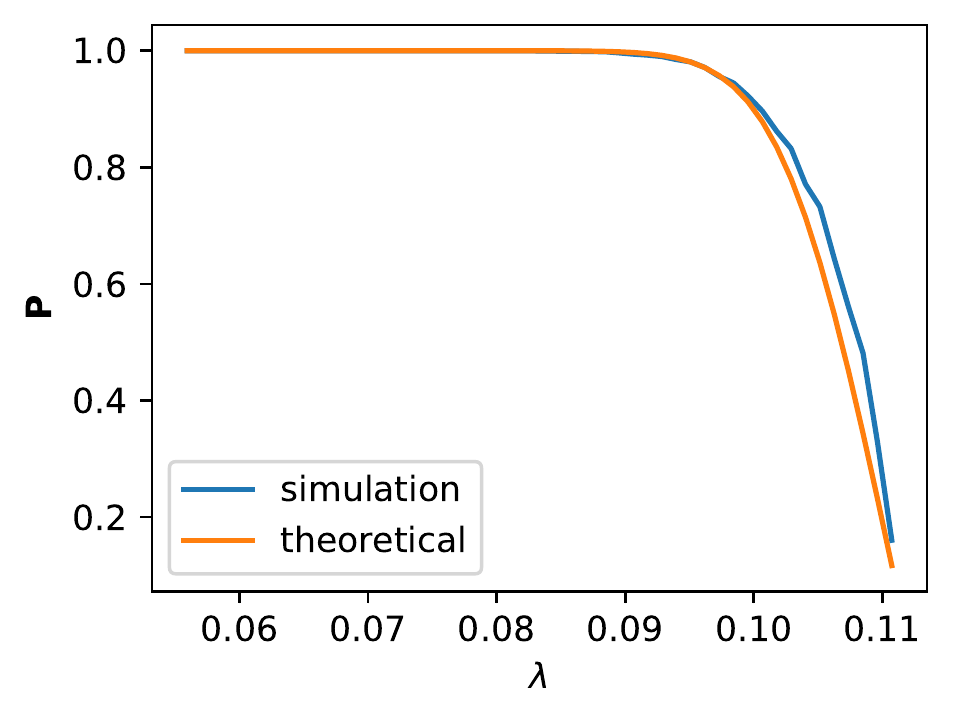}}
\caption{ Comparison of the theoretical and simulation results for the
  clearing probability $\mathbf{P}$ for a single elevator with capacity
  50. }
\label{fig:compare}
\end{figure}

We also replace the steady-state occupancy distribution $Q(N)$ by the
infinite-capacity expression in Eq.~\eqref{PQ}.  We further approximate
$Q(N)$ by a Gaussian distribution with mean $N_{ss}$ and variance Var$(N)$
given in Eq.~\eqref{varTN}; this form compares well with our simulation
results in Fig.~\ref{fig:T-distribution}.  Thus the summation in
\eqref{mathP} represents the cumulative of the convolution of an exponential
and Gaussian distribution.  This operation gives rise to the exponentially
modified Gaussian distribution~\cite{emgd}.  Thus we have
\begin{align*}
  \sum_{W}  R(W)\,\mathcal{Q}(C-W|{\rm not\,cleared}) = \Psi(u,0,v)-e^{-u+v^2/2+\log[\Psi(u,v^2,v)]}\,,
\end{align*}
where the right-hand side is the cumulative of the exponentially modified
Gaussian distribution~\cite{emgd}, and $\Psi(u,0,v)$ is the cumulative
Gaussian distribution itself, with arguments
$u=2(\lambda_c/\lambda-1)(C-\lambda T)$, mean value $0$, and standard deviation
$v=2\sqrt{\lambda T}(\lambda_c/\lambda-1)$.

Thus, Eq.~\eqref{mathP}, which determines the clearing probability
$\mathbf{P}$ becomes
\begin{align}
\label{bP}
  \mathbf{P}=\mathbf{P}\,\Psi(C,N_{ss},\sqrt{{\rm Var}(N_{ss})})
  +(1-\mathbf{P})\left\{\Psi(u,0,v)-e^{-u+v^2/2+\log[\Psi(u,v^2,v)]}\right\}\,,
\end{align}
from which $\mathbf{P}$ can be immediately obtained.  The resulting
prediction for $\mathbf{P}$ closely matches our simulations shown in
Fig.~\ref{fig:compare}.

\begin{figure}[ht]
  \center{
    \subfigure[]{\includegraphics[width=0.47\textwidth]{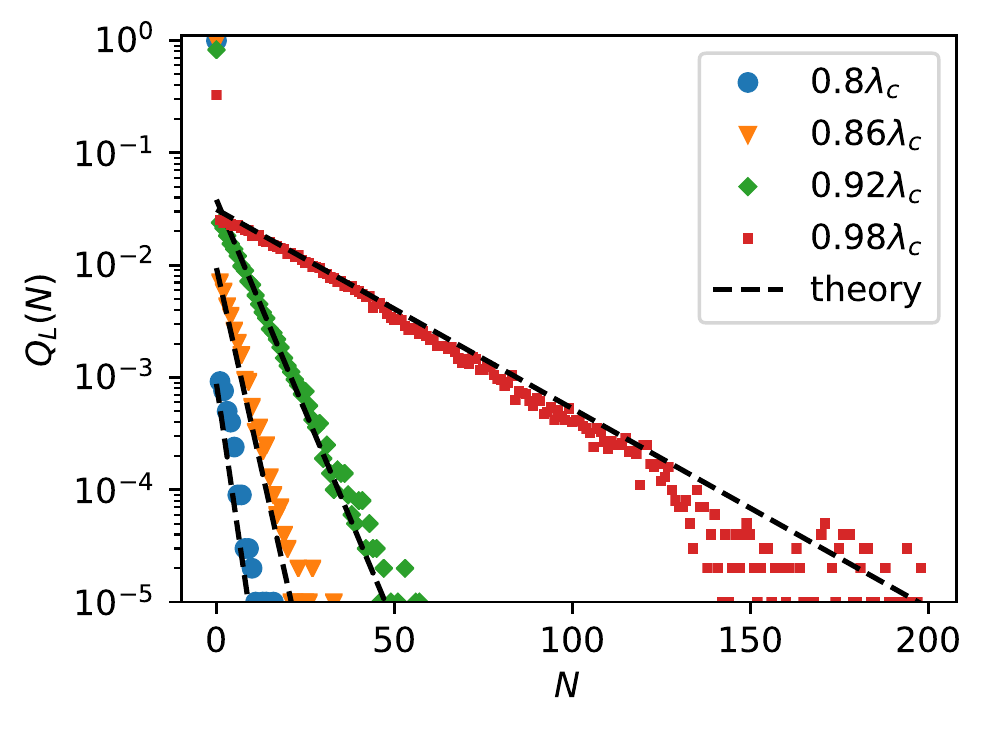}}\quad
    \subfigure[]{\includegraphics[width=0.45\textwidth]{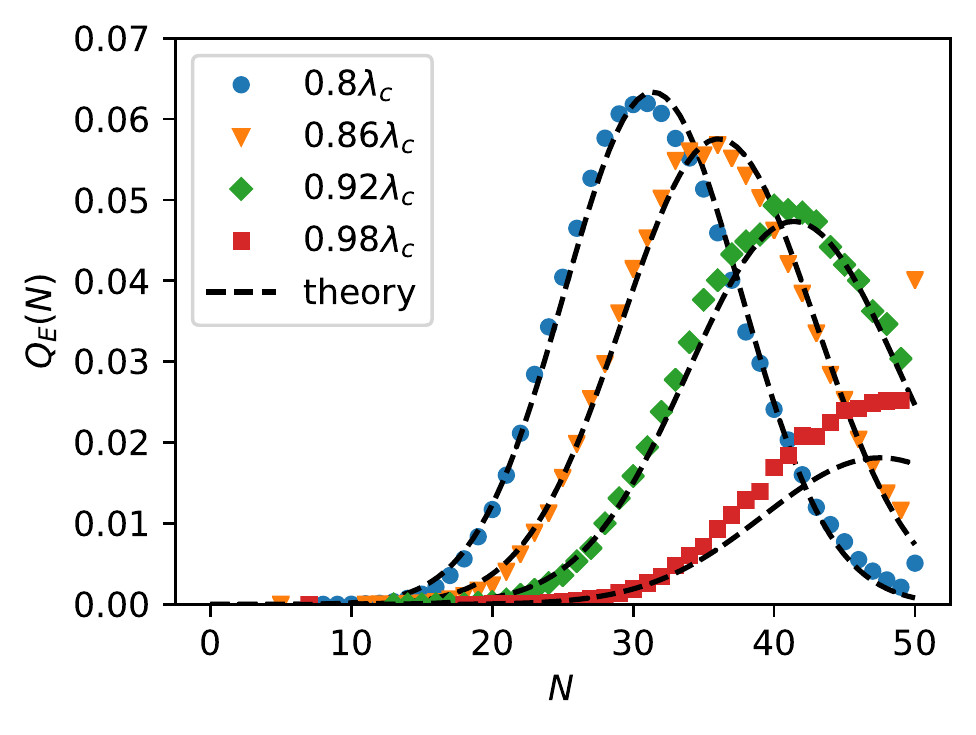}}}
   \caption{The distributions of (a) the number passengers still waiting in
     lobby, $Q_L(N)$, and (b) the number of passengers transported by
     elevator, $Q_E(N)$.  Both cases correspond to a single elevator with
     capacity $C=50$.}
\label{fig:QEQL}
\end{figure}

For a finite capacity elevator, we can now determine both the number of
passengers waiting in the lobby when the elevator departs and the number
carried away by the elevator.  Let $Q_L(N)$ and $Q_E(N)$ denote the
distributions for these two quantities. If the lobby is cleared, there will
be no passengers waiting, so that
\begin{subequations}
\begin{align}
\label{QL}
  Q_L(N)=\mathbf{P}\,\delta_{0,N}+2(1-\mathbf{P})\,(\lambda_c/\lambda-1)e^{-2(\lambda_c/\lambda-1)N}\,.
\end{align}  
If the lobby is not cleared, there will be necessarily $C$ passengers
inside the elevator, so that
\begin{align}
  \label{QE}
  Q_E(N)=\mathbf{P}\,Q(N|N<C)+(1-\mathbf{P})\,\delta_{C,N}\,,
\end{align}
\end{subequations}
where 
\begin{align*}
Q(N|N<C)=
	\begin{cases}
   {\displaystyle {Q(N)}\Big/\;{ \sum_{k=0}^{C}Q(k)}}&\qquad \text{if } N<C\,,\\[4mm]
    0              & \qquad \text{otherwise} \,,
	\end{cases}
\end{align*}
with $Q(N)$ the occupancy distribution in an infinite elevator system, which
we can well approximate by a Gaussian distribution.  Both these predictions
match our theoretical expectations, as shown in Fig.~\ref{fig:QEQL}.

We can now determine the distribution of times that an individual has to wait
until an elevator arrives.  We again need to treat two distinct cases.  If an
individual is accommodated within a single cycle, the time (s)he needs to
wait will be the same as that in the infinite capacity limit; that is,
$\mathcal{W}(t)$ in Eq.~\eqref{waitT}.  If the person needs to wait for more
then a cycle, her/his waiting time depends both on when (s)he arrives and the
number of people already waiting for the elevator.  The former attribute
determines the waiting time within a cycle whereas the latter determines how
many cycles must elapse before the person can be accommodated.  Since the
former time is uniformly distributed in $[0,T]$ for a fixed cycle time $T$,
the total waiting time will be uniformly distributed in $[mT,(m+1)T]$, with
$m$ a positive integer that is determined by the condition that the number of
people $N$ already waiting in the lobby is in range of $[mC,(m+1)T]$.
Equivalently, the number of people remaining in the lobby after an elevator
departs must be in $[(m-1)C, mT]$.

Taking into account these two cases, the distribution of waiting times,
$\mathcal{W}_C(t)$, for an elevator of capacity $C$, is formally given by
\begin{subequations}
\label{WWC}
\begin{align}
  \label{WC}
  \mathcal{W}_C(t)=\mathbf{P}_{\mathcal{W}}\mathcal{W}(t)
  +\Theta(t-T)\,(1-\mathbf{P}_{\mathcal{W}})\;
  \frac{1}{T}\; \int_{(m-1)C}^{mC} Q_L(N)\, dN\, .
\end{align}
Here $\mathbf{P}_{\mathcal{W}}$ is the probability that an individual waits
less than a single cycle time, $\Theta(\cdot)$ denotes the Heaviside step
function, $m=\text{floor}(t/T)$, $\mathcal{W}(t)$ is the waiting time
distribution \eqref{waitT} for an infinite-capacity elevator, and $Q_L(N)$ is
the distribution of the number of people waiting in the lobby (given by
Eq.~\eqref{QL}) when the elevator leaves, and we have integrated $Q_L(N)$
over the allowable range of $N$.

To make \eqref{WC} explicit, we need $\mathbf{P}_{\mathcal{W}}$.  Naively,
one might anticipate that $\mathbf{P}_{\mathcal{W}}$ should be the same as
clearing probability $\mathbf{P}$.  However, we need to account for another
possibility: even when the lobby is not cleared in the current cycle, if the
number $N$ of stranded passengers from previous cycle is less than $C$, then
$C-N$ out of the $C$ passengers that are transported by the elevator in the
current cycle will wait less than a single cycle.  By including the
additional term that accounts for this situation, the final result for
$\mathbf{P}_{\mathcal{W}}$ is
\begin{align}
  \mathbf{P}_{\mathcal{W}}=\mathbf{P}+\int _1^{C}\frac{C-N}{C}\;  Q_L(N)\,dN
  =\mathbf{P}+ (1-\mathbf{P})\,\frac{2C(\lambda_c/\lambda-1)
  -e^{-2(\lambda_c/\lambda-1) C }+1}{2C(1-\lambda_c/\lambda)}\,,
\end{align}
\end{subequations}
where we have substituted in the expression \eqref{QL} for $Q_L(N)$.
\begin{figure}[ht]
  \center{
    \subfigure[]{\includegraphics[width=0.47\textwidth]{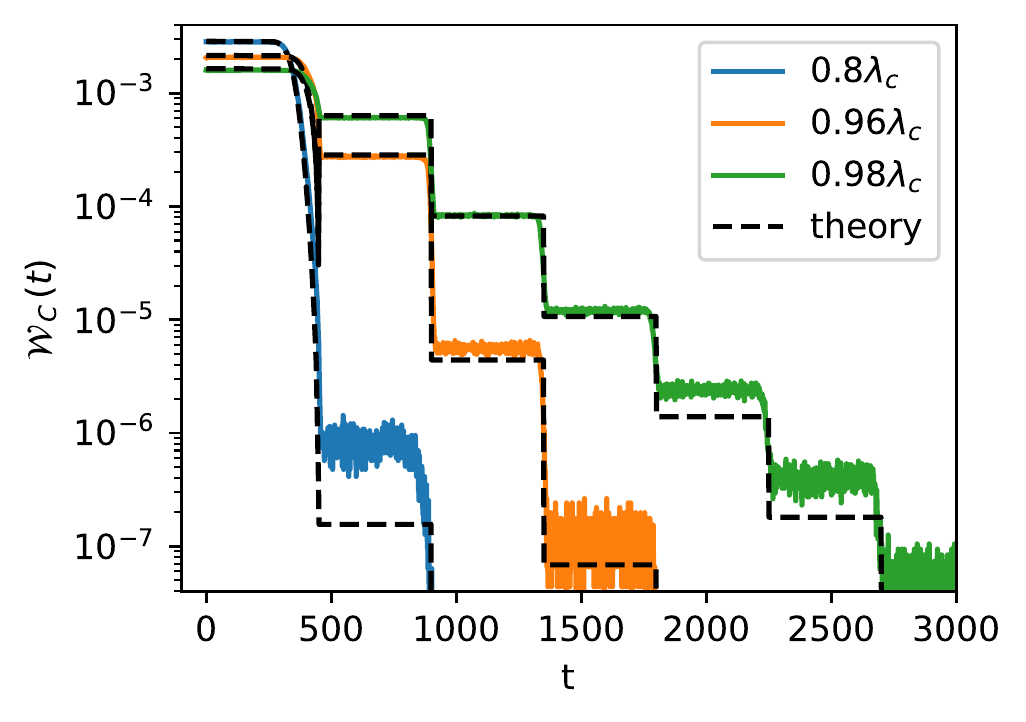}}\quad
    \subfigure[]{\includegraphics[width=0.45\textwidth]{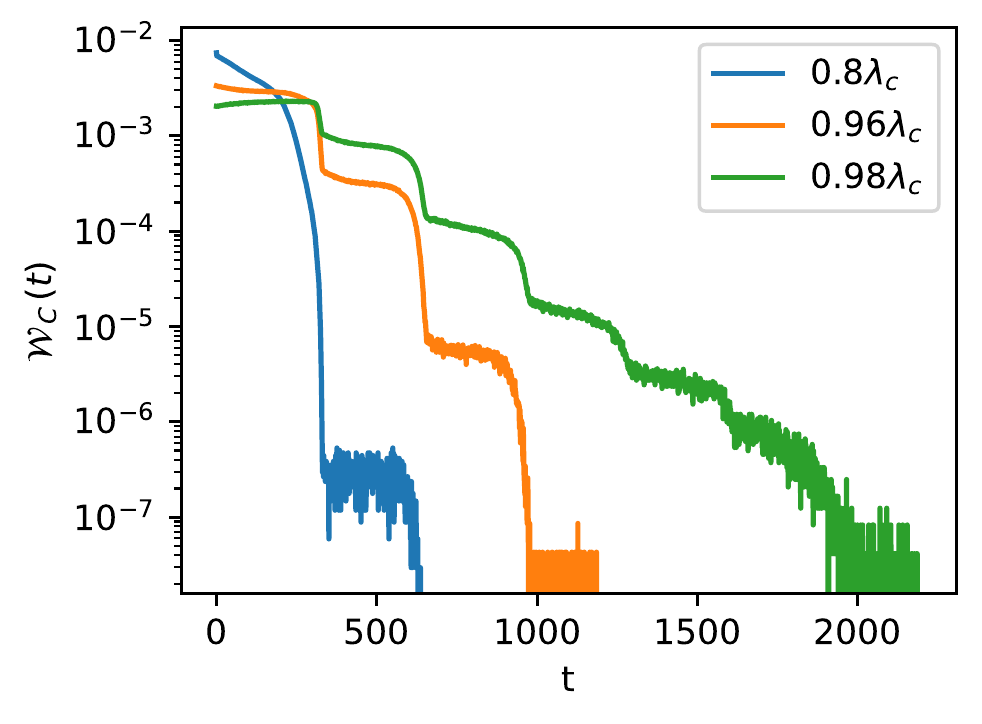}}}
  \caption{The distribution of times $\mathcal{W}_C(t)$ that a passenger
    waits until being able to board the elevator for: (a) for a single
    elevator of capacity $C=50$, and (b) for two elevators of capacity
    $C=25$. }
\label{fig:waitingT}
\end{figure}

Within each cycle, the waiting time distribution is Fermi-Dirac like, and
multiple copies of these Fermi-Dirac-like distributions constitute the full
waiting time distribution $\mathcal{W}_C(t)$
(Fig.~\ref{fig:waitingT}(a)). Qualitatively, $\mathcal{W}_C(t)$ follows an
overall exponential decay with time with a substructure that consists of
these Fermi-Dirac steps.  Since the elevator must be full if the lobby is not
cleared, these steps beyond the first are much sharper than the first one.
Figure~\ref{fig:waitingT}(a) shows a close correspondence between our
prediction \eqref{WWC} and the simulation data.

\section{Multiple Finite-Capacity Elevators}
\label{sec:k}

Finally, we now turn to the realistic situation of a building that contains
$k$ identical elevators, each of capacity $C$.  We investigate two basic
characteristics of the transport dynamics---the steady state and
synchronization.

\subsection{The steady state}
\label{subsec:ssk}

Under the assumption that the elevators are uncorrelated, the steady-state
criterion derived in Sec.~\ref{subsec:ss1} now becomes $\lambda T(C)\leq kC$.
Again using Eq.~$\eqref{TN}$ for $T(C)$, the steady-state condition is
\begin{align}
\label{lambdac-n}
  \lambda \leq \lambda_c= \frac{k}{2\tau_s+2F/(C+1)}\,.
\end{align}
For elevators with capacity $C=20$ and using $\tau_s=2.5\,$sec, $F=100$, the
steady-state criterion now gives $\lambda \leq 0.0689\,k$.  As long as the
elevators are uncorrelated, the overall transport capacity is simply
proportional to the number of elevators.

It is instructive to estimate the number of elevators of capacity $C=20$ that
are needed to service the start-of-day ``rush'' in an office building of 100
floors without having a pileup of passengers waiting in the lobby.  We assume
that each floor accommodates 100 people, so that $10^4$ people need to reach
their offices at the start of a workday\footnote{As a check of this estimate,
  the number of workers in the Willis (formerly Sears) tower in Chicago is
  roughly 15,000~\cite{willis}}.  With 20 people in each elevator, 500
elevator trips are needed.  Each trip takes roughly 5 minutes for a total of
2500 elevator minutes.  If the morning rush spans a two-hour period, these
2500 elevator minutes have to fit in a 120-minute window, which requires 21
elevators.  These 21 elevators can accommodate a total passenger arrival rate
of $\Lambda_c= 21\,\lambda_c\approx 1.5$ passengers arriving per second and
still remain in the steady state.  This total arrival rate over a two-hour
period also corresponds to accommodating all $10^4$ occupants of the
building.  If the elevators are uncorrelated, the time interval between
successive events where an elevator reaches the ground floor should equal the
single-elevator cycle time divided by the number of elevators, which is
roughly 15 seconds.  These numbers accord with common experience.

\subsection{Synchronization}

As passenger demand increases, it is not uncommon for a set of elevators to
synchronize.  This leads to the annoying feature that a large number of
passengers build up in the lobby and then multiple elevators return to the
lobby at nearly the same time.  This clustering is analogous to what occurs
in the bus-route model~\cite{o1998jamming}.  In this latter example, a
circular bus route is serviced by multiple buses, with passengers arriving at
a fixed rate at a set of bus stops.  If a bus spends a time longer than usual
at a stop because more passengers than usual are either loading or
disembarking, the following bus will tend to catch up.  Consequently, there
is less time for passengers to accumulate at stops after the leading bus has
departed.  Since the number passengers waiting at stops will be less than
average for the trailing bus, it will continue to catch up to the leading
bus.  If the trailing bus is allowed to pass the leading bus, the same
instability arises in which there are fewer passengers than average waiting
at each stop for the new trailing bus.  Overall, this instability leads to an
effective attraction between buses that tends to reduce their separation.

A similar instability occurs in a multi-elevator building when the arrival
rate of passengers in the lobby is sufficiently large.  Although each
elevator runs on its own independent track, so that elevators can effectively
``pass'' each other, the same effective attraction between elevators occurs,
which leads to the clustering of waiting passengers waiting in the lobby.  To
quantify this synchronization, let us first treat the much simpler case of
deterministic dynamics and suppose that all elevators are already
synchronized.  Because of the deterministic dynamics, the number of waiting
passengers in the lobby when all the elevators reach the ground floor equals
$\lambda$ times the cycle time.  We also suppose that $F_{\rm max}$ is
deterministic and equal to its average value of $NF/(N+1)$, where $N$ is the
number of passengers in each elevator (which is the same for each elevator).
If the number of waiting passengers is large enough to trigger the movement
of all $k$ elevators, then these elevators will again return at the same time
in the next cycle.  In turn, there will be a sufficient number of new waiting
passengers to trigger the next cycle and lock the system in a synchronized
state.  Thus within deterministic dynamics, synchronization is locked in once
it is achieved.

To trigger all $k$ elevators in the building, the number of waiting
passengers should be greater than the capacity of $k-1$ elevators.  That is,
\begin{align*}
  \lambda T(N) > (k-1)C\,.
\end{align*}
In this deterministic picture, the cycle time for all the elevators is
$T(N)= 2F_{\rm max} + 2N\tau_s$, with $N=\eta C$ and $\eta$ in the range
$(1-\frac{1}{k})$ to 1.  At the lower limit case, $\lambda$ is just large
enough that the number of waiting passengers, $\lambda T(N)$, just exceeds
the capacity of $k-1$ elevators.  At the upper limit, the number of waiting
passengers completely fills all elevators.  For simplicity we take $\eta=1$
henceforth.  Using the expression $\lambda_cT=kC$, the lower bound for
synchronization becomes
\begin{align}
  \label{sync}
  \lambda >\lambda_c\,(k-1)/k\,.
\end{align}
According to this deterministic picture, synchronization occurs in a narrow
window of arrival rates that lies between $\lambda_c(k-1)/k$ and $\lambda_c$.
This prediction is only approximate because we have not accounted for the
randomness in the cycle times of each elevator.

\begin{figure}[ht]
\center{
\includegraphics[width=18cm]{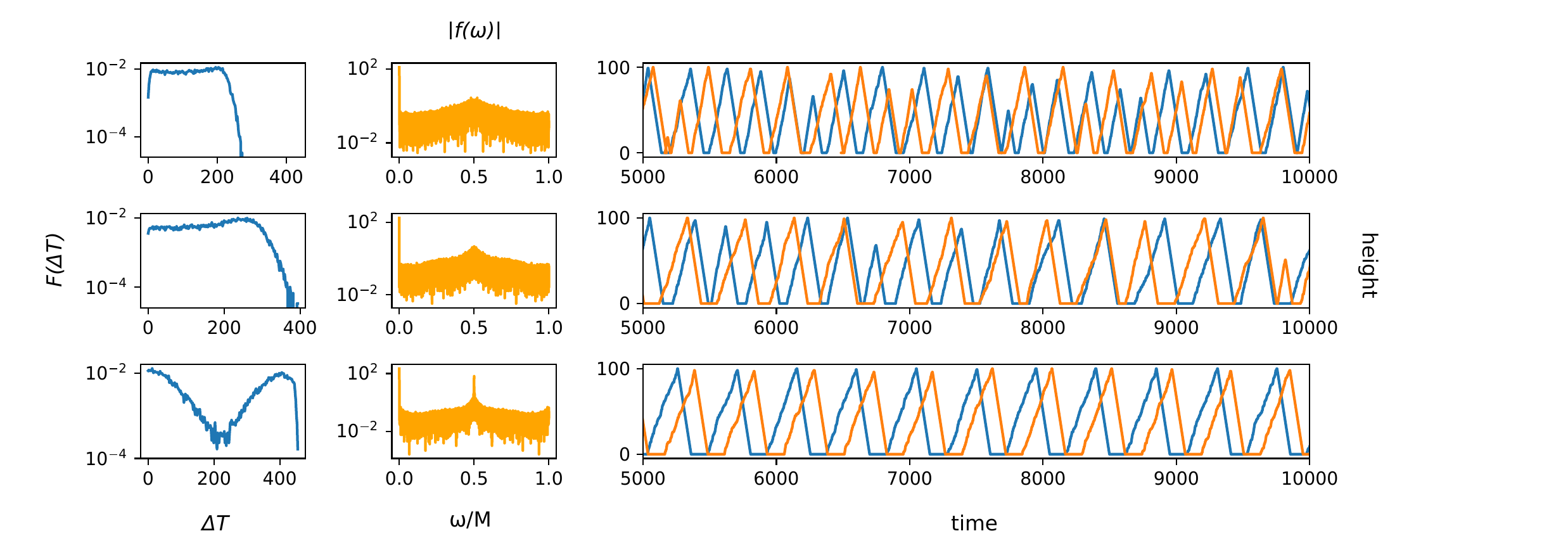}}
\caption{The inter-arrival time distribution $F(\Delta T)$ versus $\Delta T$
  (left panels), the discrete Fourier transform $f(\omega)$ of the
  inter-arrival time series versus the normalized frequency $\omega/M$,
  where $M$ is the total number of intervals (middle panels), and
  representative late-time plots of the vertical positions of two elevators
  versus time for $\lambda = 0.5\lambda_c, 0.8\lambda_c$, and $0.98\lambda_c$
  for elevators with $C=50$}
\label{fig:sync2}
\end{figure}

\begin{figure}[ht]
\center{
\includegraphics[width=15cm]{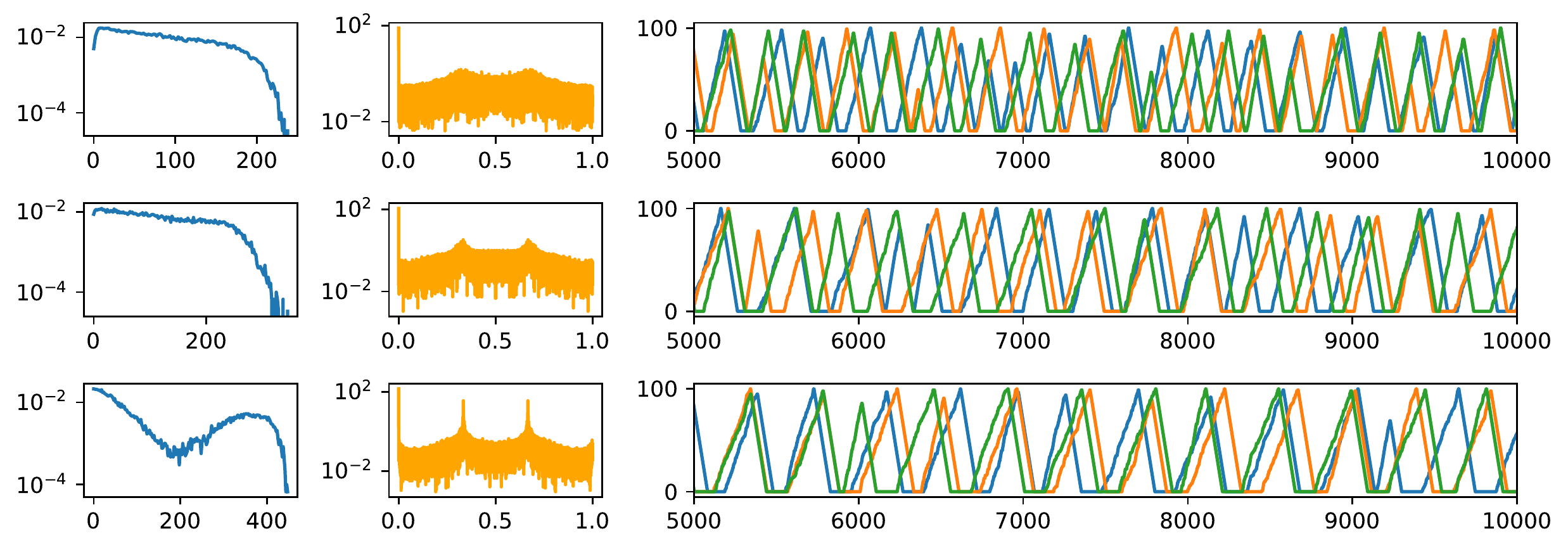}}
\caption{The same as in Fig.~\ref{fig:sync2} for a three-elevator
  system.}
\label{fig:sync3}
\end{figure}

\begin{figure}[ht]
\center{
\includegraphics[width=15cm]{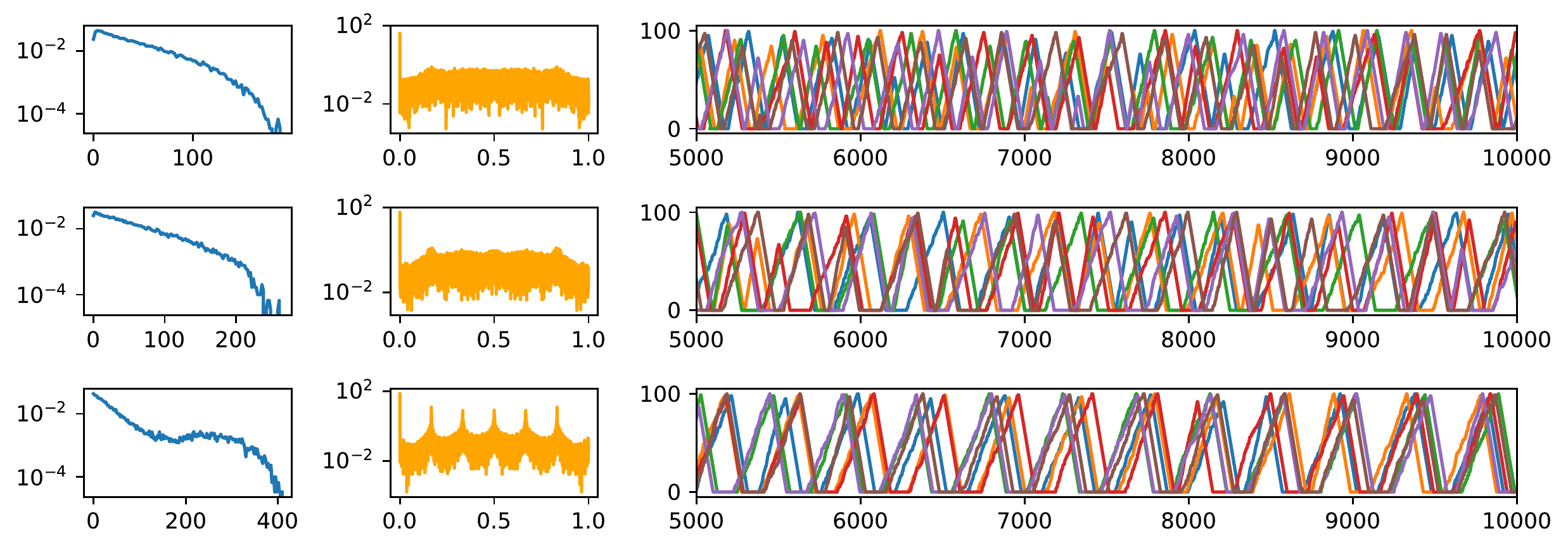}}
\caption{The same  as in Fig.~\ref{fig:sync2} for a six-elevator
  system. }
\label{fig:sync6}
\end{figure}

Illustrations of the vertical positions of each elevator versus time for a
two-elevator, three-elevator, and six-elevator system are shown in the right
panels of Figs.~\ref{fig:sync2}--\ref{fig:sync6}.  Each row consists of data
for the same value of $\lambda$.  For these cases, the elevator trajectories
tend to cluster when $\lambda\gtwid 0.9\lambda_c$.  For
$\lambda=0.8\lambda_c$, there are hints of synchronization for $k=2$ and
$k=3$, but not for $k=6$.  This behavior is expected from the deterministic
picture of Eq.~\eqref{sync}, in which synchronization requires $k\leq 5$ when
$\lambda=0.8\lambda_c$.

A useful characteristic of the many-elevator system is the time $\Delta T$
between successive elevator arrivals.  These inter-arrival times play an
analogous role to successive cycle times in the single elevator system.  When
elevators become synchronized, there should be a repetitive pattern of a long
time interval whose value is close to the single elevator cycle time,
followed by $k-1$ short time intervals that correspond to the subsequent
arrivals of nearly synchronized elevators (left panel of each figure).
Because of the near periodicity of the elevators, it is helpful to study the
Fourier transform of the ordered inter-arrival time series,
\begin{align*}
  f(\omega) = \sum_{m=0}^{M-1} \Delta T_m\; e^{-2\pi im\omega /M}\,,
\end{align*}
where $\Delta T_m$ is the $m^{\rm th}$ inter-arrival time and $M$ is the
total number of time intervals in the dataset.  In the synchronized state,
this discrete Fourier transform should have $k-1$ peaks in addition to the
component at zero frequency; this feature is illustrated in the middle panels
of each figure.

Finally, we investigate the waiting time distribution for multiple elevators.
We focus on the simplest case of the waiting-time distribution two elevators,
and simulation results are shown in Fig.~\ref{fig:waitingT}(b).  The
step-like form of the waiting-time distribution is a result of the
synchronization of the two elevators.  When the elevators become
synchronized, we can treat one cycle of the two-elevator cluster as a single
time unit for the biased random walk picture for the dynamics of number of
passengers in the lobby.  This leads to the long-time exponential decay of
the waiting-time distribution, as in the single-elevator case.  For times
that are less than a single cycle, the waiting time distribution can be again
deduced using \eqref{waitT} after replacing the cycle-time distribution
$P(T)$ by the inter-arrival times distribution $F(\Delta T)$.

\section{Concluding Comments}
\label{sec:concl}

We investigated the transport of people by elevators in a tall building
during the start-of-the-day operation within the framework of a minimal
probabilistic model.  For a single infinite-capacity elevator, we computed
the expected time for one elevator cycle and the condition for a steady-state
to arise.  In this steady state, we determined the distribution of the
elevator cycle times and occupancy distribution of the elevator.  We
constructed a rapidly converging iterative procedure that determined these
distributions.  The resulting distributions have well-defined peaks; thus
fluctuations about the average are noticeable but do not overwhelm the
dynamics.

For a single finite-capacity elevator, a new aspect of the dynamics is the
clearing time, defined as the time interval between two successive events
where all waiting passengers are accommodated by the elevator.  We argued
that this clearing time can be determined by invoking a random-walk picture
for the number of passengers that remain in the lobby when the elevator
departs.  From this picture, we computed the average clearing time and its
distribution.  We also determined the distribution of the number of
passengers who are stranded in the lobby when the elevator leaves and the
number of passengers in the elevator.  We then turned to the realistic
situation of a fixed number of finite-capacity elevators.  Again, we
determined the condition for the steady state.  Finally, we investigated the
conditions under which a set of elevators will synchronize.

Our naive approach represents a small step in developing a physical
understanding elevator transport.  There are also many natural
generalizations of the basic model that are worth considering within a our
physics-based perspective.  Perhaps the simplest is to study the situation
where the cross-sectional area $A(h)$ of the building decreases with height
$h$.  A logical question now is: does there exist an optimal profile for
$A(h)$ that minimizes the waiting time, but also optimizes available office
space?  Another relevant issue to investigate is that of elevator staging;
that is, some fraction of the elevators services floors 0 through $F/2$ and
another fraction services floors $F/2$ through $F$.  Is this configuration
better---in that the average waiting time and/or average total travel time is
shorter---than two elevators that service all floors?

It should also be worthwhile to study the case of end-of-day operation, when
occupants are leaving the building and nobody is entering.  That is,
passengers start their floor of occupancy and call for an elevator to take
them to the ground floor.  This case is not merely the time-reversed version
of start-of-day operation and it would be interesting to understand the
difference between these two cases.
  
ZFs Undergraduate Research Experience at the Santa Fe Institute was funded
by the General Collaboration Agreement for the ASU-SFI Center for Biosocial
Complex Systems.  SRs research was partially supported by NSF grant
DMR-1910736.

\appendix

\section{Event-driven simulation algorithm}
\label{app:alg}

Our numerical results are based on an event-driven simulational approach, in
which we only monitor the (variable) time interval for the next elevator to
reach the ground floor.  We describe our algorithm below, first for a single
elevator of capacity $C$, and then for multiple elevators, each of the same
capacity $C$.

\subsection*{ One elevator}

\begin{itemize}

\item If no passengers are waiting when an empty elevator reaches lobby,
  advance the time by an exponentially distributed random number $\delta t$
  whose average value equals to the inter-arrival time between successive
  passengers.  We then set $N=1$, since one new passenger is in the lobby.
  
\item If $N\geq 1$ passengers are waiting when an elevator arrives, then:
  
\begin{enumerate}

\item The waiting passengers enter the elevator until either all $N<C$ are
  accommodated or the elevator is full.  The number of waiting passengers is
  decreased by $\text{min}(N, C)$.
  
\item From the passenger destination floors, which are all chosen
  independently from a uniform distribution in $[1,F]$, determine the cycle
  time $T$ by setting $T= 2F_{\rm max}\, + 2N\tau_s$, where $F_{\rm max}$ is
  the maximum destination floor among the $N$ passengers and $\tau_s$ is
  randomly chosen from any well-behaved (i.e., no long tails) continuous
  distribution with mean value 2.5.  In our simulations, we use a uniform
  distribution.

\item Increment the time by $T$ and populate the lobby with $N$ new
  passengers, with $N$ chosen from a Poisson distribution with mean value
  $\lambda T$.  The passengers currently in the elevator are erased.
  
\item The elevator picks up the waiting passengers and a new cycle begins.

\end{enumerate}

\end{itemize}

\subsection*{$k>1$ elevators}

We need to now track the return time of each of the $k$ elevators and an
event is defined by the return of the elevator with the shortest current
return time.

\begin{itemize}

\item If no passengers are waiting when an empty elevator reaches the ground
  floor, advance the time by a random number $\delta t$ whose average value
  equals is the inter-arrival time between successive passengers, and then
  set $N=1$.  Decrement the return times of all other elevators by
  $\delta t$.

\item If $N\geq 1$ passengers are waiting when an elevator arrives, then:
  
\begin{enumerate}

\item Passengers enter the elevator until either all $N<C$ passengers are
  accommodated or the elevator is full.  The number of waiting passengers is
  decreased by $\text{min}(N, C)$.  
  
\item From the set of passenger destination floors, determine the return time
  $T$.

\item Increment the time by $T_{\rm min}\equiv \text{min}\{T_i\}$, where
  $T_i$ is the current travel time of the $i^{\rm th}$ elevator to reach the
  lobby and $i$ runs from $1$ to $k$.  Decrement the travel times of all
  other elevators by $T_{\rm min}$.  Populate the lobby with $N$ new
  passengers, with $N$ chosen from a Poisson distribution with mean value
  $\lambda T_{\rm min}$.
  
\item The elevator with the minimum return time picks up the waiting
  passengers and a new cycle begins.

\end{enumerate}

Note that loading an elevator, elevator travel, and elevator unloading are
combined into a single event in this algorithm.  Consequently, there is no
possibility for a second elevator to arrive in the lobby during the time that
one is currently loading passengers.

\end{itemize}

\bigskip\bigskip

%\bibliographystyle{iopart-num}
%\bibliography{references.bib}

\providecommand{\newblock}{}

\end{document}